\documentclass[letterpaper]{article}
\usepackage{aaai2026}
\usepackage{times}
\usepackage{helvet}
\usepackage{courier}
\usepackage[hyphens]{url}
\usepackage{graphicx}
\usepackage{natbib}
\usepackage{caption}
\usepackage{booktabs}     
\usepackage{multirow}      
\usepackage{threeparttable} 
\usepackage{array}

\providecommand{\answerTODO}[1]{#1}
\providecommand{\TODO}[1]{}
\providecommand{\todo}[1]{}

\title{The Digital Landscape of God: Narrative, Visuals and Viewer Engagement of Religious Videos on YouTube}
\begin{document}

\author{
  Rongyi Chen\textsuperscript{\rm 1}\thanks{This work was completed during Rongyi Chen’s visiting research at George Mason University.},
  Ziyan Xin\textsuperscript{\rm 2}\equalcontrib,
  Qing Xiao\textsuperscript{\rm 3}\equalcontrib,
  Ruiwei Xiao\textsuperscript{\rm 3},
  Jingjia Xiao\textsuperscript{\rm 4},
  Bingbing Zhang\textsuperscript{\rm 5},
  Hong Shen\textsuperscript{\rm 3},
  Zhicong Lu\textsuperscript{\rm 6}
}

\affiliations{
  \textsuperscript{\rm 1}Peking University, School of Journalism and Communication, Beijing, China\\
  \textsuperscript{\rm 2}Neuroscience Institute, Carnegie Mellon University, Pennsylvania, United States\\
  \textsuperscript{\rm 3}Human-Computer Interaction Institute, Carnegie Mellon University, Pittsburgh, Pennsylvania, United States\\
  \textsuperscript{\rm 4}University of California San Diego, San Diego, California, United States\\
  \textsuperscript{\rm 5}School of Journalism and Mass Communication, University of Iowa, Iowa City, Iowa, United States\\
  \textsuperscript{\rm 6}Department of Computer Science, George Mason University, Fairfax, Virginia, United States\\
  qingx@andrew.cmu.edu
}
\maketitle

\begin{abstract}
The digital transformation of religious practice has reshaped how billions of people engage with spiritual content, with video-sharing platforms becoming central to contemporary religious communication. Yet current research lacks a systematic understanding of how narrative and visual elements create meaningful spiritual experiences and foster viewer engagement. We present a mixed-methods study of popular religious videos on YouTube across major religions, developing taxonomies of narrative frameworks, visual elements, and viewer interaction. Using LLM-assisted analysis, we studied relationships between content characteristics and viewer responses. Findings show that religious videos predominantly adopt speaking-style formats with authority-based persuasion strategies, using salvation narratives for guidance. All prefer bright lighting, with Buddhism favoring warm tones and prominent symbols, Judaism preferring indoor settings, and Hinduism emphasizing sacred objects. We identified differentiated patterns of emotional sharing among religious viewers while revealing significant correlations between content characteristics and engagement, particularly regarding AI-generated content.
\end{abstract}

\section{Introduction}

Religion and spiritual belief are core elements that profoundly shape the lifestyles and social interactions of billions worldwide. Over the past decades, digital media have fundamentally transformed religious content access and community participation~\cite{claisse2023keeping, campbell2024looking}. The COVID-19 pandemic accelerated this shift, as global religious communities' acceptance of remote services has surged dramatically, with hybrid online-offline models becoming standard practice for many religious groups~\cite{przywara2021online, claisse2023keeping, wolf2022spirituality}.

Large numbers of believers are engaging with and deepening their faith through watching religious video tutorials, online sermons, and interactive religious educational content, with online video dominating digital religious practices~\cite{campbell2021digital}. The audiovisual format can vividly convey abstract theology, complex spiritual experiences, and emotional resonance. 46\% of Americans have used at least one online religious technology product, and 20\% watch religious-themed videos from video-sharing platforms such as YouTube and TikTok~\cite{faverio2023online}. Specifically, 58\% of historically Black Protestants, 47\% of Evangelicals, 28\% of mainline Protestants, 24\% of Catholics, and 19\% of Jews watch at least once monthly~\cite{faverio2023online}. Views of videos tagged \texttt{\#religion} on TikTok surged from 2 million to over 2 billion within two years~\cite{Tait2021TwoBillion}. This content not only sustains traditional communities but also brings ultimate concern questions into public discourse through innovative narrative strategies, attracting new members and expanding their influence~\cite{mosavel2022religiosity}.

Despite this substantial expansion, a systematic understanding of how content characteristics in popular religious videos relate to viewer engagement remains limited. Specifically, existing literature lacks comparative, multimodal, and cross-religious analyses conducted on YouTube. Few studies have empirically connected specific audiovisual elements to engagement outcomes or examined the emerging influence of AI-generated content on spiritual communication. While online communities have been extensively studied, the mechanisms of collaborative meaning-making and identity formation in religious discourse spaces remain underexplored. To address these gaps, we propose four research questions: \textbf{(RQ1a)} What narrative frameworks do popular religious videos employ to construct their content? \textbf{(RQ1b)} What visual elements characterize religious video production across traditions? \textbf{(RQ2)} What discourse patterns emerge in religious video comment sections as spaces for collective meaning-making? \textbf{(RQ3)} How do content characteristics correlate with different dimensions of user engagement?

To address these questions, we employ a mixed-methods approach combining LLM-assisted multimodal coding with human validation and large-scale computational analysis. We specifically focus on the ecosystem of popular religious content on YouTube across five major traditions: Christianity, Islam, Hinduism, Buddhism, and Judaism. Our dataset comprises 1,100 high-engagement videos, spanning both long-form and short-form formats, and includes 1.92 million associated user comments. Beyond mere descriptive statistics, we treat this data as rich archival material containing personal reflection, emotional resonance, and spiritual experience, and utilize statistical models to rigorously examine the correlations between audiovisual production choices and patterns of viewer engagement.

For \textbf{RQ1a}, we employed LLM-assisted coding with manual validation by two human coders to systematically annotate narrative frameworks, including video types, speech strategies, religious narratives, and persuasion strategies. Our analysis revealed a core persuasive template centered on salvation narratives and authority-based appeals. For \textbf{RQ1b}, we analyzed visual elements including scene types, lighting, color tones, religious symbols, and production techniques, uncovering distinct visual aesthetics across faiths, such as warm tones in Buddhist videos and dramatic lighting in Christian content. For \textbf{RQ2}, we applied fine-tuned LLM-assisted coding validated through human review to analyze user discourse patterns, examining thematic discussions and emotional dimensions. This uncovered a dual emotional structure where expressions of transcendent awe and approval coexist with significant presence of anger and sarcasm. For \textbf{RQ3}, we employed chi-square tests and correlation analysis to examine relationships between content characteristics and engagement outcomes.

This study makes two contributions. First, we offer a large-scale, cross-religious analysis of religious videos on YouTube, linking narrative and visual features to patterns of viewer engagement and comment discourse. Second, we demonstrate how religious narratives and visual aesthetics shape collective emotional expression and processes of meaning-making in online faith communities.

\section{Related Work}

This research draws from two interrelated domains: (1) The Digital Ecosystem of Mediatized Religion, and (2) Visual and Narrative Design in Religious Media and its relationship with user engagement. By reviewing these domains, we aim to frame the relationship between content characteristics and community engagement patterns in online religious spaces.

\subsection{The Digital Ecosystem of Mediatized Religion}

Religious and spiritual individuals have long used digital means to stay connected and engage in communal practices such as prayer, worship, or meditation \cite{golan2019religious}. In this ecosystem, religious communities are not passive recipients of technological influence; they actively adopt and adapt digital media tools to achieve their goals of spiritual dissemination and community maintenance on a broader scale \cite{przywara2021online,  hutchings2011contemporary}. Audiences, in turn, increasingly rely on these platforms for continuous exchange and learning with other members of their religious groups, especially when their own mobility is constrained \cite{claisse2023keeping, wolf2022spirituality, yin2026active}.

Video has emerged as a key medium, fundamentally altering the structures of religious authority, the pathways of information dissemination, and the modes of community interaction \cite{hoover2016rethinking}. Video platforms like YouTube and TikTok provide not only large-scale, low-cost distribution channels but also create unprecedented mechanisms for audience engagement through algorithmic recommendations, interactive features, and community-building tools \cite{przywara2021online, al2022social, xiao2025institutionalizing}. Digital platforms have empowered users to transcend geographical boundaries, granting them access to a global array of denominations, theological perspectives, and cultural expressions. This has prompted a shift in religious practices from traditional “inherited belonging” to active “individualized seeking” \cite{anderson2023you, xiao2026enhancing}.

\subsection{Visual and Narrative Design in Religious Media}

The core challenge in digitizing religious content lies in translating abstract theological concepts and transcendent spiritual experiences into tangible and affectively resonant elements \cite{claisse2023keeping, wolf2022spirituality}. Visual language plays a key role in creating aesthetic experiences and user engagement \cite{becker2017audiovisual}. Digital religious content extensively leverages visual design principles to convey spiritual meaning \cite{markum2024mediating}, and through visual effects, can create a sense of virtual presence that surpasses live experiences \cite{golan2019religious, lim2020live}. Moreover, AI-driven video generation and virtual studio technologies are enabling a democratization of content creation, allowing lay believers and small groups to produce religious content with richer visual appeal \cite{campbell2017religious}.

Religious videos function as persuasive media. Images are not mere reproductions of reality but persuasive tools that through symbols and composition construct arguments and guide emotions~\cite{messaris1997visual}. Prior research on YouTube vlogs has suggested that visual presentation strongly influences viewer perception ~\cite{biel2011known}. Their core function is to reproduce the multisensory experience of sacred physical spaces online \cite{lim2020live}, echoing the theory of mediatized ritual where media become sites of ritual performance \cite{sumiala2014mediatized}. It is through meticulously designed audiovisual aesthetics, including lighting, color palettes, music, and camera movements, that these videos cultivate an atmosphere evoking spiritual experiences among viewers \cite{bekkering2019studying, mansour2022holy}.

YouTube comment sections serve as crucial sites for user engagement, functioning as sociotechnical systems for identity performance, social support, and fostering belonging \cite{hutchings2011contemporary, kraut2020makes}. These spaces exhibit a complex discursive ecology where users engage in theological debates and affective expressions \cite{grigoropoulou2020discussing}, often revealing group biases and polarization \cite{al2015online}. Campbell's theory of “digital religion” highlights four key levels of negotiation in online religious communities: authority, identity, authenticity, and community \cite{campbell2020digital}. Different visual narrative strategies can guide user attention and emotional expression. However, how multimodal content characteristics in religious videos systematically relate to collective discourse patterns remains underexplored.

\section{Data \& Methods}
This study employs a mixed-methods approach combining large-scale data collection with AI-assisted multimodal analysis to examine religious video content and user engagement. Our methodology is structured around three analyses: content framework analysis using automated coding of audiovisual elements, discourse pattern analysis via comment classification, and statistical correlation analysis connecting content characteristics with engagement outcomes.

\subsection{Data Collection}
Our research focuses on popular religious videos from YouTube. We selected adherent names for five major world religions as search keywords: Christianity, Islam, Hinduism, Buddhism, and Judaism. Following prior work~\cite{maram2024topic, ccelebiouglu2022muslim}, this keyword-based approach captures videos containing religious content regardless of whether the search term appears explicitly in the title, as YouTube indexes videos based on comprehensive metadata including descriptions, tags, and transcripts.

We developed an automated Python script using the YouTube Data API v3, intentionally employing the API rather than browser-based collection to minimize the influence of personalized recommendation algorithms. Sorting by view count, we retrieved metadata for the top 120 videos for each keyword from both standard videos and YouTube Shorts sections, initially acquiring 1,200 records. We selected view count as the sorting criterion because it serves as a relatively objective indicator of content reach and ensures substantial comment volumes for discourse analysis. Our English-language queries returned multilingual results due to YouTube's cross-language indexing.

We conducted manual curation to refine the dataset, excluding: (1) content containing explicit terrorism or extremism, due to incompatibility with our LLM analysis pipeline's safety filters; and (2) scripture-chanting music videos with static images, due to lack of dynamic visuals and sparse comments. After filtering, a final corpus of 1,100 videos was established, with an average viewership of 5.06 million views. The 1,100 videos originated from 820 unique creators across 49 countries, with the United States (N=208, 27.80\%) and India (N=188, 22.93\%) most represented. Detailed statistics are provided in the appendix.

For RQ1, we downloaded video source files preserving original resolution and performed LLM-assisted coding with human validation. For RQ2, we collected 1,924,477 comments distributed across Christianity (N=576,222), Islam (N=446,226), Judaism (N=367,376), Hinduism (N=343,390), and Buddhism (N=191,263). The comment corpus spans over 130 languages, with English as the predominant language (76.04\%), followed by French (2.01\%), Chinese (1.73\%), Spanish (1.64\%), and German (1.44\%). After removing non-substantive text, hyperlinks, and garbled characters, 1,920,132 comments remained for analysis. For RQ3, we constructed statistical models to explore relationships between video features and user responses. This study was approved by the Institutional Review Board (IRB) at the first author's institution.

\subsection{Method: Analysis of Video Narrative Frameworks (RQ1a)}

To systematically explore presentation patterns of religious videos at scale, we implemented a rigorous three-stage pipeline: (1) inductive codebook development from 100 pilot videos; (2) automated coding using Qwen3 via few-shot prompting; and (3) comprehensive human verification. We employed large language models as auxiliary coding tools, an emerging methodology for enhancing analytical efficiency in qualitative research~\cite{li2025text, zhang2025dark}. Given current limitations in multimodal recognition reliability, we define our approach as LLM-assisted coding. Crucially, no labels utilized in the final analysis were purely model-produced. Two human coders carefully reviewed and validated all model outputs to ensure accuracy, aligning with recent findings that LLMs achieve optimal performance when combined with strict human oversight~\cite{khalil2025evaluating, dunivin2025scaling}.

We developed a multimodal analysis framework comprising two parallel analytical paths: analysis of audio transcriptions through a Narrative Framework, and systematic coding of the visual stream through a Visual Framework. We define a video's content framework as the creator's core expressive intent conveyed through speech. We first classified videos based on overall presentation format (narration, music, dance, ritual recording, or other forms), then conducted standardized audio extraction.

Two researchers developed a content analysis codebook through inductive manual coding of a subset of video samples, drawing on prior research on video narrative frameworks~\cite{montero2020digital, zhang2023understanding} and religious narrative strategies~\cite{yassif2016storytelling, roof1993religion}. We established three core analytical dimensions: (1) Rhetorical Mode, identifying the macro-structure and presentation form; (2) Religious Narrative Framework, focusing on the worldview framework embedded in the video; (3) Persuasion Strategy, analyzing rhetorical devices employed to enhance influence.

We deployed Qwen3 (qwen3:32b) in a local server environment for large-scale coding. Qwen3-32B was selected for its comprehensive coverage of over 100 languages and competitive performance on general capability benchmarks, where it achieves results comparable to frontier models such as Grok-3 and Gemini-2.5-Pro. Local deployment was adopted to prevent transmission of potentially sensitive user-generated data to external servers, thereby safeguarding participant privacy. We utilized few-shot learning; each prompt included the codebook definitions, three to five human-coded examples per category, and the target transcript segment. The model was instructed to output a single category label accompanied by a brief rationale, which human coders subsequently reviewed during the validation stage. Prior to full-scale coding, 100 videos were independently coded by two human coders and the model, achieving substantial inter-coder reliability (Krippendorff's $\alpha$=0.86, indicating acceptable agreement). Audio transcription was also performed using Qwen-series models, with all transcribed texts stored in plain text format for subsequent analysis. The automated coding pipeline incorporated a multi-round validation mechanism: the model coded each item independently across three separate runs, and any item yielding at least one inconsistent result across the three runs was flagged and submitted to human coders for final adjudication.

\subsection{Method: Analysis of Video Visual Elements (RQ1b)}

The visual framework analysis examines how videos use audiovisual elements to cultivate sanctity and spiritual experience. We referenced classic theories of visual grammar from social semiotics~\cite{harrison2003visual}, treating visual elements as a symbolic system with specific cultural meanings. Two researchers conducted open, iterative visual coding on 100 videos, incorporating dimensions from previous studies~\cite{lim2020live, wolf2022spirituality}: color~\cite{bekkering2019studying}, lighting~\cite{mansour2022holy}, location~\cite{kong2001religion}, AI-generated content~\cite{simmerlein2025sacred}, B-roll techniques~\cite{huber2019b}, religious symbols~\cite{xue2024faith}, and character attire~\cite{schmidt1989church}. The codebook distinguishes between speaker-focused visual elements and supplementary material elements.

For preprocessing, we standardized all videos to 480p resolution and capped analysis at the first ten minutes. This protocol aligns with the model's official input specifications to mitigate context overflow risks and ensure inference stability, while retaining sufficient visual fidelity for semantic recognition. Pilot coding of a subset of long-form videos indicated that primary narrative and visual cues, including opening rhetorical strategies and dominant visual aesthetics, are predominantly established within this window, providing sufficient representativeness for the dimensions analyzed in this study. For visual coding, we used Qwen-VL (qwen-vl-max-2025-04-08), a flagship multimodal LLM from the Qwen family, to assist in identifying key frames and content distribution patterns. The LLM analyzed entire video files directly, capturing temporal information such as camera movement and scene transitions. The prompt explicitly instructed the model to annotate its outputs according to video timestamps, thereby preserving temporal information for downstream analysis. Crucially, the LLM served as an auxiliary identification tool, with two human coders manually reviewing and validating all visual coding results to ensure accuracy. We achieved inter-coder reliability of Krippendorff's $\alpha$=0.83, indicating acceptable agreement for visual content analysis.

\subsection{Method: Analysis of Comment Discourse (RQ2)}

We employed LLM-assisted coding with human validation to capture complex semantics, affects, and interactional intentions within comments. The codebook was constructed through a hybrid iterative process: we first conducted LDA topic modeling on English-language comments to identify thematic patterns, then two researchers performed open coding on a subset to identify emergent patterns. These categories were refined using established frameworks from prior research on interactional intent~\cite{claisse2023keeping} and online religious communities~\cite{rotman2010wetube}.

We finalized three analytical dimensions: (1) Expressive Paradigm, categorizing interactional intent (expressing stance, questioning, etc.); (2) Comment Theme, identifying 11 core topics; (3) Affective Expression, quantifying 8 religiously-relevant affects including awe and tranquility. The Expressive Paradigm dimension was coded at two levels of granularity: a broader level (L1) capturing general interactional intent such as stance-taking, questioning, and sharing, and a refined level (L2) identifying more specific motivations within each L1 category, such as agreement, opposition, and experiential testimony.

We selected Qwen-32B, a flagship model known for strong multilingual understanding, deployed locally with specialized fine-tuning using Low-Rank Adaptation (LoRA). We constructed a gold standard dataset of 5,000 comments independently coded by two researchers (Krippendorff's $\alpha$=0.91, indicating high agreement), which served as the fine-tuning corpus. During inference, each prompt included the codebook definitions, three to five gold-standard examples per category, and the target comment; the model was instructed to output a single category label with a brief rationale. The fine-tuned model significantly outperformed the base model. To assess generalization on unseen data, we validated model predictions on a held-out random sample of 1,000 comments against independent human coders, achieving Krippendorff's $\alpha$=0.84, indicating acceptable agreement for comment classification.

\subsection{Method: Statistical Association Analysis (RQ3)}

We conducted Chi-squared ($\chi^2$) tests to explore associations between video features (RQ1) and comment patterns (RQ2). Given the large sample size, we prioritized effect size (Cram\'{e}r's V) over simple significance testing to identify meaningful patterns rather than trivial statistical differences. We performed Cochran's rule checks and applied Bonferroni correction for multiple comparisons to ensure robustness. We also implemented stratified analyses based on video length and religious affiliation to identify moderating effects. This exploratory approach aims to uncover potential relationships between production choices and engagement outcomes, providing directional insights rather than causal claims.

\section{Findings: Narrative Frameworks (RQ1a)}
\label{RQ1A}

\subsection{Video Type}

\begin{figure}[tb]
  \centering
  \includegraphics[width=\columnwidth]{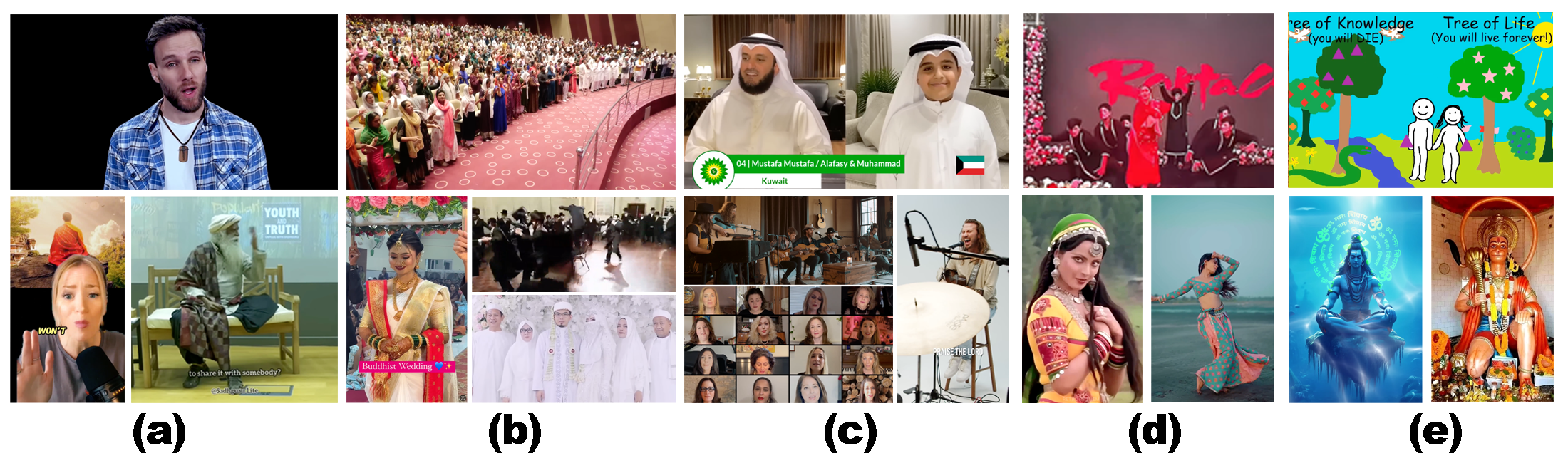}
  \caption{Examples of video types: (a) Speaking-style, (b) Event recording, (c) Music, (d) Dance, and (e) Other.} 
  \label{fig:001} 
\end{figure}

Speaking-style videos are the primary format (62.09\%), featuring individuals facing the camera to deliver content explanations, covering topics such as religious worldviews and commentary on controversial events (Figure~\ref{fig:001}(a)). These videos frequently document sermons and extensively employ B-roll editing techniques. Buddhist videos demonstrate the highest prevalence at 70.32\%.

On-site documentation of religious rituals or events comprises 17.27\% of content, including collective activities like chanting and choir performances, as well as individual ceremonies such as weddings (Figure~\ref{fig:001}(b)). Music videos (12.45\%) show significant increases in normal videos, with Christian and Jewish music being particularly active. Dance occupies a minimal proportion, while other formats include animated short films and AI-generated visual presentations of religious deities (Figure~\ref{fig:001}(e)).

\subsection{Speech Strategy}

Religious videos demonstrate diverse communication approaches. \textbf{Teaching \& Guidance} (34.73\%) involves instructing spiritual concepts and practices, such as step-by-step explanations of the “Noble Eightfold Path” in Buddhist videos or verse-by-verse analysis of biblical passages. Buddhist videos show the highest prevalence (47.49\%). \textbf{Preaching \& Proclamation} (9.00\%) refers to formal religious speeches by religious leaders in religious settings, such as pastors delivering Sunday sermons or imams giving Friday sermons. Christianity shows higher usage (16.52\%), while this strategy decreases in short videos (-6.76\%) as effective preaching requires deeper context.

\textbf{Interview \& Dialogue} (11.18\%) involves interactive formats like Q\&A sessions and roundtable discussions, evenly distributed across religions. Islamic videos show prominence (16.11\%), typically featuring scholars discussing jurisprudential issues. \textbf{Debate \& Apologetics} (6.45\%) involves defending faith viewpoints or refuting dissenting arguments, such as debates on evolution versus religious creation. This strategy is significantly higher in short videos (+7.79\%), likely due to algorithms may favor polarization. Christianity demonstrates higher proportions (11.61\%), while Buddhism uses this less frequently (2.28\%), consistent with teachings prioritizing compassionate dialogue~\cite{sangma2024spiritual}.

\textbf{Testimony \& Sharing} (4.09\%) involves personal stories of faith conversion and spiritual practice, such as “How I came to faith” or “How Buddhist teachings changed my life,” characterized by high personalization and emotional authenticity. \textbf{Stories \& Parables} (10.91\%) involves retelling religious classics and moral fables. Hinduism and Buddhism show more activity due to rich mythological traditions~\cite{appleton2016shared}.

\subsection{Religious Narrative Framework}
\label{RQ1A_3}

\textbf{Salvation Narrative} dominates religious videos (46.36\%), emphasizing themes of error, repentance, and redemption. Christian videos show the highest proportion (63.84\%), typically depicting arcs from sin through inner struggle to spiritual liberation. A typical example: a sharer recounts being “bound by alcohol and anger” until pastoral guidance led to transformation. This narrative depth corresponds with a significant reduction in short videos (-14.35\%). Hindu videos show greater diversity with lower salvation proportions (34.84\%), as Hindu practice encompasses multiple paths including karma yoga and devotional yoga.

\textbf{Progress Narrative} (9.55\%) emphasizes positive change and self-improvement, showing prominence in Buddhism (20.55\%). Buddhist progress narratives revolve around gradual practice from ignorance to enlightenment~\cite{marques2012consciousness}, such as staged progress in meditation and compassion. \textbf{Harmony Narrative} (8.36\%) emphasizes peace and unity among different groups, increasing in short videos (+3.60\%). Buddhism shows prominence (15.98\%), a pattern that may align with its cultural emphasis on compassion and the Middle Way~\cite{sangma2024spiritual, kraft1992inner}. \textbf{Conflict Narrative} (5.64\%) emphasizes confrontation with opposing values, more frequent in short videos (+7.62\%) as confrontational content is typically associated with high engagement.

\subsection{Persuasion Strategy Framework}
\label{RQ1A_4}

\textbf{Authority Dependence} strategies dominate religious videos (42.27\%), reflecting the central role of classic texts and religious leaders in religious communication. Christian videos extensively quote biblical verses, Islamic videos cite Quran and Hadith, and Buddhist videos reference Buddha's teachings. Islamic videos show most prominence (47.87\%), often displaying Arabic scripture with translations while an imam explains its application to daily life. This strategy significantly decreases in short videos (-10.52\%), as authority citations are compressed into soundbites that may weaken persuasive depth. \textbf{Experiential Evidence} strategies (34.36\%) use personal experiences to persuade, significantly increasing in short videos (+8.97\%). Creators share faith conversions, life improvements from religious practice, and miraculous experiences. A pilgrim displays the journey from the Sichuan-Tibet line to Lhasa, arriving at the Potala Palace with tears: “Now I understand gratitude and resilience more than when I started.” The preference for authenticity on short video platforms makes them ideal for this strategy.

\textbf{Logical Argument} strategies (5.36\%) use rational reasoning to support religious viewpoints, such as cosmological arguments or design arguments to prove God's existence. A video blogger reasons at a whiteboard: “Anything with a beginning must have a cause. The universe has a beginning. Therefore, the universe must have a 'first cause' transcending space and time.” \textbf{Emotional Resonance} strategies (3.36\%) appeal to emotion through music, visual arts, and direct emotional expressions. Buddhism shows the highest proportion (7.76\%). \textbf{Scientific Support} strategies (1.45\%) cite scientific research to support religious viewpoints, such as neuroscience research on meditation's brain benefits. Hinduism shows relative prominence (3.62\%), likely due to scientific attention to yoga and meditation~\cite{tang2015neuroscience}. \textbf{Identity Identification} strategies (0.73\%) emphasize group identity and belonging. Judaism shows prominence (2.67\%), a pattern that may be consistent with its ethno-religious identity and documented traditions of group cohesion~\cite{katz2024can, sosis2004adaptive}, through content emphasizing shared history and cultural traditions.

Full distributions of comment coding results across all religions and video formats are reported in the appendix.

\section{Findings: Visual Elements (RQ1b)}
\label{RQ1B}

\subsection{Scene Type}
Scene selection is the key to constructing atmosphere and a sense of place. Outdoor scenes (22.64\%) encompass natural environments, urban spaces, and historical sites, associated with connotations of openness and universality. Personal studios (11.82\%) emulate professional media production with controlled lighting; Hindu videos show higher usage (16.29\%), possibly related to yoga and meditation instruction requirements.

Religious venues including churches, temples, and mosques comprise only 7.55\%, establishing sacred identity and reinforcing credibility. Islamic and Buddhist videos show higher usage (10.90\% and 10.96\% respectively). Other indoor venues (31.36\%) reflect religious communication's extension into everyday spaces such as homes and conference halls, dissolving boundaries between sacred and secular. Jewish videos show the most prominent indoor usage (43.56\%), creating a life-oriented, demystified effect. Solid/virtual backgrounds (12.45\%) show particular prominence in Buddhist videos (25.57\%), used to visualize Pure Land scenes or meditation states.

\subsection{Shot Types}

Shot types control rhythm, emotional distance, and emphasis. Close-up shots (18.18\%) amplify emotion and create intimacy; Buddhist videos demonstrate the highest usage (28.77\%). Medium shots (31.18\%) balance facial expression with environmental context, employed significantly more in short videos (37.75\% vs 24.59\%) to maximize visibility in limited screen space.

Wide shots (4.54\%) establish environment and atmosphere. Mixed shots (36.36\%) use variation to maintain rhythm and attention; normal videos employ these significantly more than shorts (48.09\% vs 24.68\%), reflecting greater need for visual variation in longer content. Islamic videos demonstrate the highest mixed shot usage (40.28\%).

\subsection{Lighting Design}

\begin{figure}[tb]
  \centering
  \includegraphics[width=0.8\columnwidth]{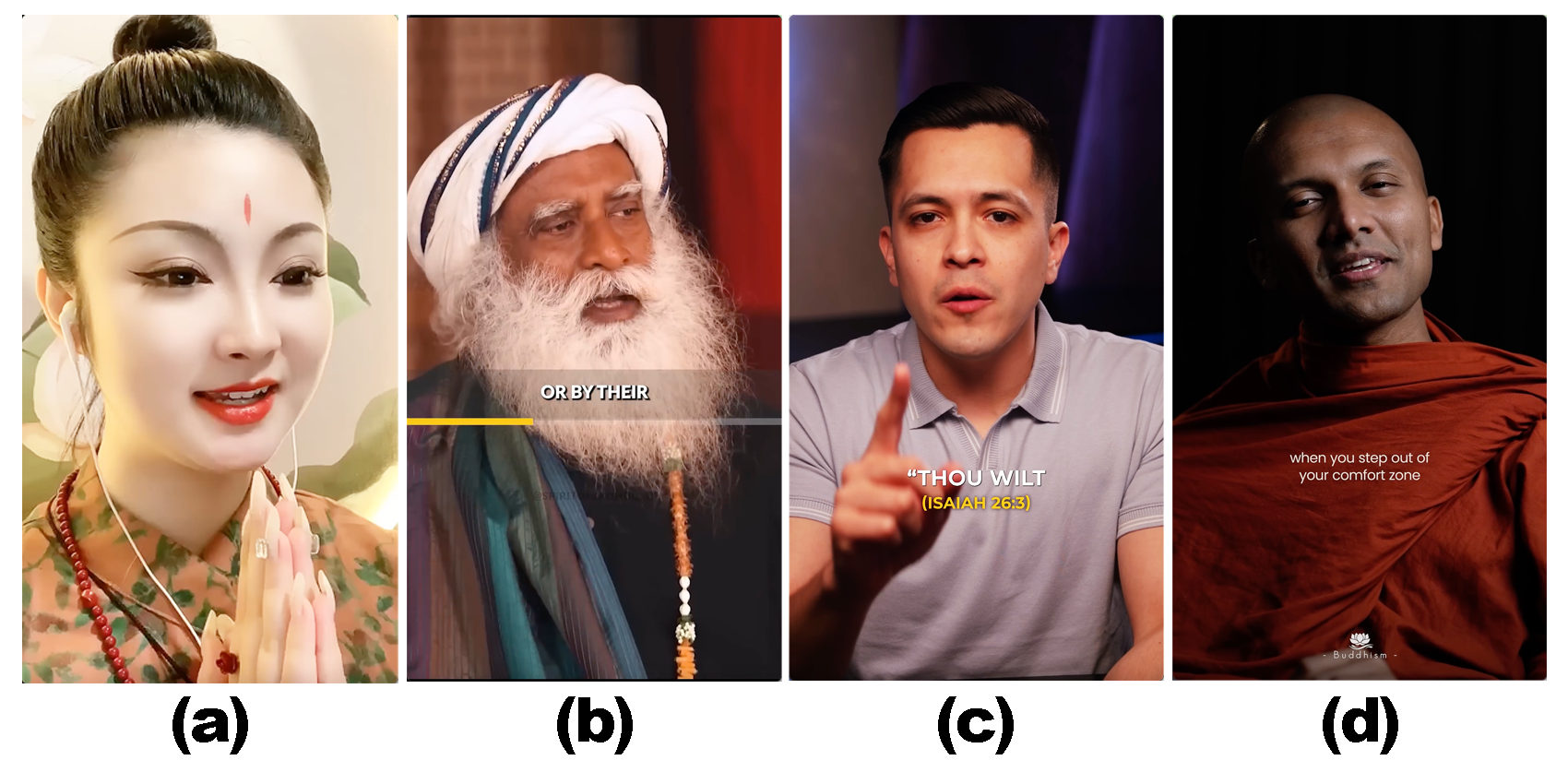}
  \caption{Examples of lighting designs in religious videos: 
(a) bright lighting, (b) natural lighting, 
(c) dramatic lighting, and (d) dim lighting.} 
  \label{fig:002} 
\end{figure}

Lighting is key for creating atmosphere and expressing symbolic meaning. Bright lighting holds absolute dominance (73.0\%), conveying positive, open, and authentic effects (Figure~\ref{fig:002}(a)); Islamic and Jewish videos show the highest usage (77.73\% and 77.33\%). Natural lighting (9.82\%) conveys authenticity and unforced natural states(Figure~\ref{fig:002}(b)), typical in morning prayer sharing or mountaintop spiritual experiences; Christian videos show higher usage (13.84\%). Dramatic lighting (4.36\%) uses strong light-shadow contrasts for symbolic and professional effects (Figure~\ref{fig:002}(c)); Christian videos show higher usage (10.27\%). Dim lighting (2.18\%) creates mysterious or solemn atmospheres (Figure~\ref{fig:002}(d)), suitable for meditation or deep contemplation content.

\subsection{Color Tone Design}
\label{RQ1B_4}

\begin{figure}[tb]
  \centering
  \includegraphics[width=\columnwidth]{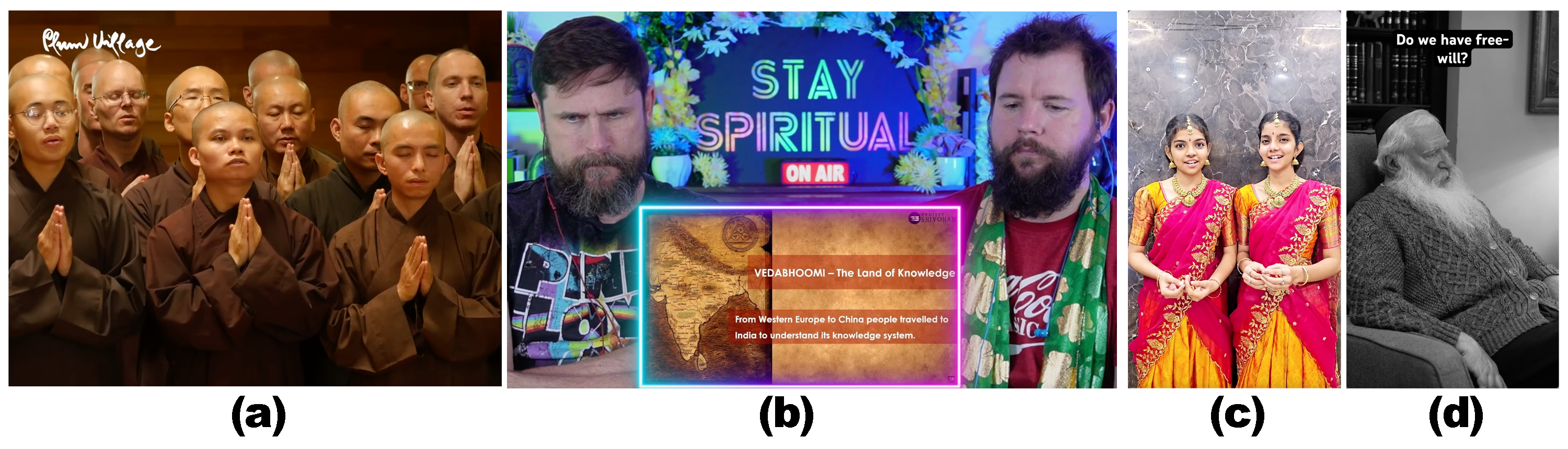}
  \caption{Examples of color tone designs in religious videos: 
(a) warm tones, (b) cool tones, 
(c) neutral tones, and (d) achromatic tones.} 
  \label{fig:003} 
\end{figure}

Color tone is another key visual element~\cite{falk2018seeing}. Warm tones primarily include red, orange, yellow, and their variations  (Figure~\ref{fig:003}(a)), with relatively longer wavelengths. Cool tones mainly include blue, green, purple, and their variations (Figure~\ref{fig:003}(b)), with relatively shorter wavelengths. Neutral tones refer to balanced color combinations with no obvious warm or cool tendencies in visual presentation (Figure~\ref{fig:003}(c)). Black, white, and gray belong to the achromatic system, lacking hue attributes and expressing visual information solely through brightness variations (Figure~\ref{fig:003}(d)). This framework of warm, cool, and achromatic tones is used primarily for emotional guidance. Religious videos overall are dominated by neutral/balanced tones (51.27\%). Short and normal videos demonstrate distinct differences in color tone usage. Short videos show a stronger preference for neutral tones (58.44\% / 44.08\%), while normal videos are more inclined to use warm tones (36.25\% / 27.95\%).

Jewish and Islamic videos show the most prominent use of neutral tones (67.56\% and 63.98\% respectively). Buddhist videos demonstrate the highest usage of warm tones (52.51\%), a pattern that may be consistent with Buddhism's cultural and artistic traditions emphasizing warmth and compassion, though regional production conventions likely also contribute. Warm tones are emotionally inviting, reducing the psychological distance with the audience. The common golden and orange-red schemes in Buddhist videos may draw on traditional artistic conventions, potentially reinforcing the warmth associated with Buddhist teachings~\cite{sangma2024spiritual, kraft1992inner}. Correspondingly, cool tones suggest rational thinking, solemn reverence, and transcendent emotions, particularly suitable for serious religious topic discussions. Christian videos show relative prominence in cool tone usage (9.82\%). Black and white tones account for 2.09\% in religious videos. Though rare, they carry symbolic meaning (e.g., classic, eternal), eliminating color distraction to highlight historical depth and solemnity~\cite{falk2018seeing}.

\subsection{Religious Symbols \& Sacred Items}

\begin{figure}[tb]
  \centering
  \includegraphics[width=\columnwidth]{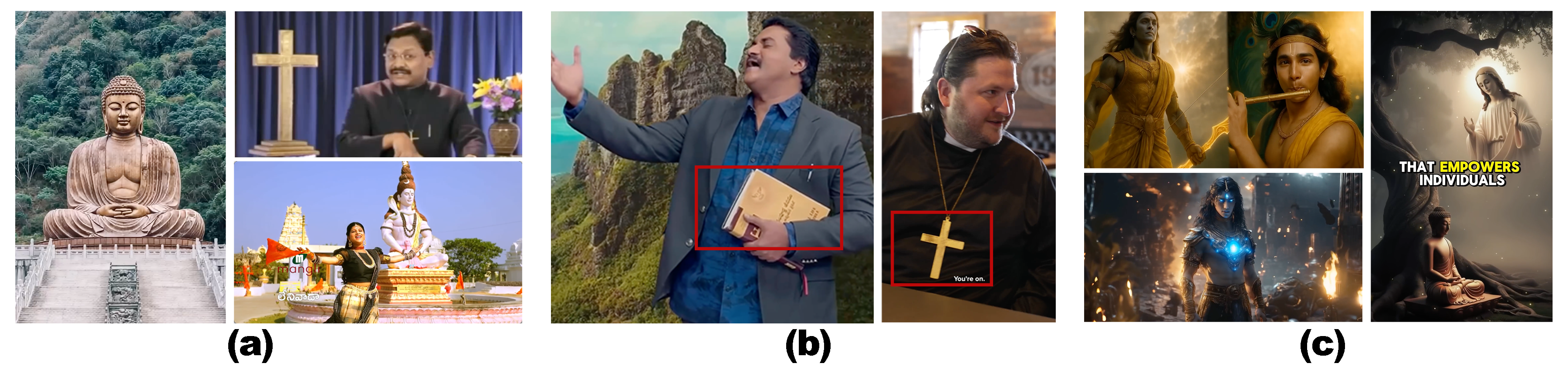}
  \caption{Examples of religious symbols and sacred items: 
(a) symbolic imagery including crosses, crescent moons, 
and dharma wheels, (b) physical sacred objects including 
scriptures, ritual implements, and devotional items, 
and (c) AI-generated visual representations of religious scenes.} 
  \label{fig:004} 
\end{figure}

Religious symbols and sacred items are key visual identifiers, serving to recognize identity, transmit culture, and express faith~\cite{bacquet2016religious}. Symbols are patterns or images with symbolic meaning (Figure~\ref{fig:004}a), while sacred items are physical objects (Figure~\ref{fig:004}b); together they form a system of religious visual expression. In symbol display, secondary symbols (39.36\%), visible but not focal, are dominant, often integrated into backgrounds or clothing for subtle identity recognition. Primary symbol display (34.91\%) strongly establishes identity and authority. Buddhist and Hindu videos show exceptional primary symbol usage (52.51\% and 51.13\% respectively), reflecting their rich visual traditions. Short videos exceed normal videos in secondary symbol usage (44.28\% vs 34.43\%), while normal videos surpass shorts in primary symbol usage (39.34\% vs 30.49\%). Typical manifestations include prominent Buddha statues, Hindu deity images, or illuminated crosses.

In sacred item display, most religious videos do not use sacred items (60.18\%), especially short videos (+2.32\%), reflecting a preference for simplicity in digital communication. The use of mixed sacred items (18.09\%) decreases significantly in short videos (-6.43\%), suggesting challenges in adapting complex combinations to short formats. Religious books account for 12.36\%, as the most common single sacred item type, symbolizing textual authority. Buddhist videos show the highest sacred item usage rate, with only 41.55\% having no sacred items, reflecting Buddhism's emphasis on material carriers and ritual implements. In contrast, Christian and Islamic videos rely less on sacred item display (no sacred items at 72.32\% and 72.99\% respectively), emphasizing textual transmission.

\subsection{AI-Generated Content}
The use of AI in religious video production represents an emerging frontier~\cite{alam2025blind, chrostowski2025verse}. We define AI-generated content as videos containing visually identifiable synthetic elements, including AI-generated imagery of religious figures, AI-animated scenes, and AI-synthesized backgrounds or visual effects. This definition excludes AI-assisted audio, scripts, or editing workflows that do not manifest as visible synthetic content. In our analysis, we categorized the applications of AI-generated elements in videos and distinguished these from traditional production methods. AI adoption remains limited but is on an upward trajectory; 5.28\% of videos incorporated AI-generated elements. This relatively low adoption rate reflects both the nascent stage of AI application in religious media production and ongoing debates surrounding authenticity, spiritual authority, and the relationship between artificial intelligence and divine inspiration. Importantly, the presence of AI-generated content should not be construed as indicative of deceptive intent or as a repudiation of religious tradition. It represents an emergent creative practice through which producers give visual form to spiritual imagination and disseminate it via digital media. Compared with conventional approaches such as hand-drawn illustration or three-dimensional animation modeling, AI-assisted production substantially lowers technical barriers and improves workflow efficiency, though its specific applications within religious communication contexts remain an open area for future inquiry. Short-form content exhibits notably higher AI adoption rates than long-form videos. Short-form creators leverage AI to overcome resource constraints, produce visually compelling material, and remain competitive on algorithm-driven platforms that reward novelty.

\subsection{Costume Design}
Casual wear dominates overall religious videos (47.36\%). This choice conveys accessibility and authenticity, reducing psychological distance. Christian videos show the most prominent use of casual wear (66.07\%). This clothing strategy visually signals the integration of religion into everyday life, lowering communication barriers. Buddhist videos show the most prominent use of religious attire (35.62\%), which may reflect monastic dress codes and the visual conventions of Buddhist media production. Typical attire includes robes, kasayas, and ceremonial hats that identify religious status and establish the speaker's credibility. These garments can convey complex information about practice level, sect, and merit. Formal wear (e.g., suits, formal dresses) comprises 7.0\% of religious videos, with relative prominence in Jewish videos (13.78\%), reflecting an identification with professional imagery and respect for the audience. It is primarily seen with religious scholars in media interviews or formal settings.

\subsection{B-Roll Integration}\label{broll}
B-roll is an editing technique that uses auxiliary footage to create visual hierarchy and rhythm~\cite{huber2019b}. In our coding process, we identified videos containing additional material beyond primary content as employing B-roll elements, distinguishing them from videos relying solely on main footage. 56.45\% of videos incorporate B-roll. This prevalence indicates the widespread adoption of professional video production techniques within religious communication. B-roll serves multiple functions: it provides visual breaks during lengthy theological discussions, illustrates abstract spiritual concepts, and maintains visual interest during audio-focused segments such as prayers or scripture recitations. Hindu videos demonstrate the highest adoption rate at 74.66\%. Christian videos have a 58.93\% adoption rate, while Buddhist, Islamic, and Jewish traditions show more moderate B-roll usage levels. Normal videos significantly exceed short videos in B-roll usage, with the 16.38 percentage point difference reflecting the distinct production approaches and audience expectations of different video formats. The longer duration of normal videos allows for complex narratives that benefit from B-roll's visual variation. 

\section{Findings: The Discourse Ecology of Comment Communities (RQ2)}
\label{RQ2}
\subsection{Basic Expressive Paradigms (Pattern L1)}
\label{RQ2_1}
Our analysis revealed distinct user expression patterns. “Stance” comments dominate (54.11\%), suggesting that articulating a position serves as the primary motivation for commenting. “Sharing” comments (15.07\%) point to a shift in traditional top-down religious communication, with ordinary users becoming active participants in discourse transmission. “Questioning” comments (9.71\%) suggest that digital platforms may provide space for rational debate even in religious contexts. The relatively low proportions of “Seeking Interaction” (1.86\%) and “Call to Action” (0.82\%) indicate a general preference for unidirectional expression.

Short videos show increased “Stance" comments (+3.67\%) while “Sharing” decreases (-3.55\%), suggesting that short formats may favor rapid emotional expression over in-depth sharing. Christian videos exhibit the highest “Stance” proportions (60.25\%), while Jewish videos are most active in “Questioning” (13.31\%), patterns that may relate to cultural characteristics of different traditions. The “Other” category (18.43\%) includes content that cannot be classified into specific patterns, such as promotional messages, simple expressions like “hmm” or emoji-only responses, and recommendation-related comments. We retain this category because such undifferentiated content remains part of the discourse ecology, even when it does not display explicit religious engagement or thematic substance.

\subsection{Refined Interactive Motivations(Pattern L2)}
\label{RQ2_2}
In refined analysis, “Agreement \& Approval” emerges as dominant (43.69\%), suggesting that audiences tend to engage with content affirming their existing values. However, “Opposition \& Denial” (7.19\%) indicates these spaces function beyond simple echo chambers; such comments increase on short videos (+4.32\%), possibly due to algorithms that favor controversial content. The relatively high opposition rate in Jewish videos (9.59\%) appears consistent with scholarly observations about traditions valuing rigorous debate as a means of exploring faith~\cite{katz2024can}.

“Experience \& Testimony” sharing (6.63\%) is particularly prevalent in Christian videos (9.46\%), a pattern that aligns with the documented role of personal testimony in Christian community building~\cite{thornton2016youtube}. This in-depth sharing sharply decreases on short videos (-4.66\%), while concise “Knowledge Sharing” increases (+1.90\%). “Factual Questioning” (5.29\%) shows higher rates in Jewish videos (7.65\%), which may resonate with the tradition's emphasis on textual literacy.

\subsection{Comment Topics}
Thematic analysis reveals dual characteristics of digital religious spaces. “Doctrine \& Creed” (19.55\%) and “Religious Figures” (17.65\%) remain prominent, suggesting the continued salience of traditional religious content in online discussions. Christian videos show the highest doctrine discussions (24.79\%), a pattern consistent with the tradition's emphasis on systematic theology. Hindu videos show lower doctrine proportions (11.26\%) but the highest “Religious Figures” discussions (26.26\%), potentially reflecting the central role of deities in Hindu devotional practice.

Notably, “Other Topics” constitute the largest proportion (32.61\%), encompassing off-topic discussions, platform-related comments, and content lacking clear thematic focus. Rather than excluding this category, we retain it to capture the authentic composition of the discourse ecology, acknowledging that not all user engagement aligns with religious or substantive themes. The presence of “Video Feedback” (9.90\%) and “Social Issues” (4.71\%) further suggests that these spaces are not insulated religious forums but rather deeply integrated with broader social concerns.

Full distributions of comment coding results across all religions and video formats are reported in the appendix.

\subsection{Emotional Expression}
\label{RQ2_4}

Our analysis reveals a complex emotional landscape with a “dual-peak structure”: “Awe” (26.12\%) and “Approval” (25.52\%) dominate, suggesting digital platforms effectively convey sacred and transcendent experiences. The higher prevalence of “Awe” in Christian videos (33.12\%) may resonate with theological emphases on divine transcendence; the prominence of "Calmness" in Buddhist videos (22.00\%) may align with meditative goals of inner tranquility; the comparatively lower “Awe” in Jewish videos (18.15\%) might reflect intellectual traditions that interweave rational inquiry with affective experience, though these patterns may equally reflect creator demographics and regional audience compositions.

However, negative emotions are also prevalent. “Anger” (9.24\%) and “Sarcasm” (7.43\%) increase significantly on short videos, suggesting platform algorithms optimizing for high-arousal content may coincide with amplified conflict. Higher negative emotions in Jewish, Hindu, and Islamic video comments may reflect real-world geopolitical controversies manifesting online. The presence of “Confusion” highlights cognitive challenges that complex religious ideas pose to audiences.

\section{Findings: Connecting Content Characteristics with User Engagement (RQ3)}
\label{RQ3}

\subsection{Associations with Cognitive Engagement}
\label{RQ3_1}

Our analysis reveals that video features correlate significantly with user expression patterns, a proxy for cognitive engagement. AI-generated content demonstrates a strong association (Cram\'{e}r's V = 0.387). Videos with moderate AI use co-occurred with more complex comment patterns, suggesting that viewers may hold ambivalent orientations toward visual enhancement in spiritual contexts. Heavy AI use corresponded with a significant increase in “Questioning” behaviors, potentially reflecting authenticity concerns.

Video type showed moderate associations (V = 0.157). Sermon/lecture videos were more frequently linked to “Call to Action” responses, while testimony videos co-occurred with higher levels of “Criticism and Sarcasm,” possibly reflecting challenges to the authenticity of personal experiences. Color tone achieves moderate association strength (V = 0.147), indicating that visual emotional orientation relates to expression choices. Notably, content features like speech strategy (V = 0.140) and narrative framework (V = 0.126) had weaker associations, suggesting technical presentation may align more closely with certain engagement patterns than narrative structure does.

At the refined level (L2), the association with AI-generated content is further pronounced (V = 0.410), indicating that technical features relate to user expression motivations in multi-layered ways. Religious narrative frameworks demonstrate enhanced associations (V = 0.191); historical narratives strongly correlate with “Experiential Testimony” sharing, while personal testimony narratives are linked to higher instances of “Criticism and Sarcasm.” Speech strategy associations are similarly enhanced (V = 0.177), with narrative strategies primarily co-occurring with “Affirmation and Approval.”

\subsection{Associations with Comment Topics}

AI-generated content showed the strongest association with topical focus (V = 0.448), the highest value across all dimensions, indicating a notably strong association with user attention allocation. Videos with moderate AI application were linked to Religious Practice discussions, while heavy application co-occurred frequently with Religious Stories topics.

Sacred object display demonstrates significant associations with topic distribution (V = 0.192). Multiple sacred objects correlate with “Religious Stories” discussions, videos without sacred objects are linked to more “Religious Practice” topics, while single objects primarily relate to “Religious Figures” discussions. This suggests that sacred objects, as visual symbols, may serve to anchor audience attention toward related content.

Video type (V = 0.136) and shot type (V = 0.132) achieve moderate associations, indicating that specific filming techniques have measurable relationships with topical attention. In contrast, narrative frameworks and speech strategies demonstrate weaker associations (V = 0.114 and V = 0.110), aligning with the pattern that technical features appear dominant in shaping topic selection.

\subsection{Associations with Emotional Engagement}

AI-generated content presents moderate-to-strong associations with emotional engagement (V = 0.210), indicating that technology intensity relates to emotional response patterns. Religious narrative frameworks demonstrate strong associations (V = 0.199), with historical narratives strongly correlating with “Awe,” suggesting that narrative structures maintain emotional guidance functions in digital environments. Speech strategy associations are similarly significant (V = 0.186), with narrative strategies primarily co-occurring with “Awe.” The consistent associations between content features and emotional expression suggest that audiovisual production choices maintain meaningful emotional relevance across digital media contexts. Sacred object display also achieves moderate associations (V = 0.149), reflecting the unique role of religious symbols in emotional resonance. Complete correlation results are reported in the appendix.

\section{Discussion}

\subsection{The Production of the Sacred}

It should be noted that cross-tradition comparisons in this study warrant cautious interpretation. Observed differences may reflect creator ecosystems, regional production conventions, and platform algorithmic dynamics as much as, or more than, the doctrinal or theological distinctions inherent to each tradition. 

Our multimodal analysis indicates that the persuasive potential of religious videos appears linked not to narrative or visual elements in isolation, but to their synergistic interplay. Specific combinations, such as warm tones with \textit{Salvation Narratives} in Buddhist videos or dramatic lighting with authority-dependent strategies in Christian content, form holistic persuasive aesthetics. This extends prior media studies by suggesting that in spiritual contexts, aesthetic choices may serve as a primary vehicle for value transmission~\cite{claisse2023keeping, wolf2022spirituality, campbell2021digital}. The prevalence of authority-dependent persuasion strategies points to a potential tension in digital religious communication. A paradox seems to emerge where the democratic affordances of platforms, which allow anyone to become a creator, interact with the hierarchical validation mechanisms central to many faith traditions. Furthermore, the widespread adoption of professional production techniques like B-roll suggests a sophisticated media literacy among religious creators, who appear to strategically use visual variation to manage cognitive load when presenting abstract theological concepts.

While AI adoption remains nascent~\cite{alam2025blind, chrostowski2025verse}, its strong association with user engagement highlights a critical challenge. The skepticism we observed toward AI-generated content, where viewers appreciated visual appeal while questioning spiritual authenticity, implies that in domains concerning core beliefs, user trust may rely more on perceived genuineness than technological sophistication. This observation contributes to broader debates about AI-mediated communication by illustrating how authenticity concerns may intensify in value-laden contexts.

\subsection{Online Community and the Performance of Faith}

YouTube comment sections function as liminal spaces that are neither purely religious nor entirely secular, but rather hybrid domains for value negotiation and identity construction~\cite{rotman2010wetube, gargi2011large}. The prevalence of stance-taking and sharing suggests a shift from passive content reception to active performance of belief, forming polycentric discourse structures. This indicates that value-laden discussions may foster more participatory, less hierarchical forms of interaction compared to task-oriented communities. The dual emotional structure, characterized by high levels of awe and approval alongside substantial anger and sarcasm, reflects the affective complexity of these spaces. Digital platforms appear to facilitate both transcendent experiences and theological conflicts, hosting emotional landscapes that are arguably more intense than traditional religious settings. Our data suggest that platforms may preserve rather than homogenize tradition-specific discourse characteristics: Jewish communities exhibited higher questioning rates, while Buddhist communities showed a prominence of calm expressions. This implies that platforms for value-based communities should support the fluid roles users adopt in co-creating meaning, rather than assuming uniform interaction patterns.

\subsection{The Algorithmic Context of Digital Spirituality}

The systematic differences between short- and long-form videos suggest that platform formats correlate with distinct shifts in religious narratives. The lower frequency of \textit{Salvation Narratives} and the higher incidence of conflict narratives in short-form videos imply that algorithmic optimization for immediate engagement may exist in tension with contemplative spiritual practices, a dynamic that co-occurs with higher prevalence of controversial content in short-form formats.

Our analysis points to a trade-off between cognitive depth and emotional intensity. Short-form videos appear effective at eliciting immediate emotional responses yet are associated with lower levels of deep cognitive engagement, such as sharing personal testimonies. This suggests that different spiritual functions may require fundamentally different technological affordances. Religious communities appear to demonstrate a sophisticated understanding of this media ecology, engaging in what looks like a tactical shift from authority-dependent strategies in long-form content to experiential evidence in short-form videos. Whether this pattern reflects deliberate strategic choice or emergent convention remains an open question, though it may broadly indicate adaptation to platform-specific affordances. This raises ethical questions for platform governance. Algorithms driven by engagement metrics tend to favor content maximizing watch-time over content nurturing spiritual growth. Our findings suggest that religious communities are developing platform-specific expressions shaped by technological constraints rather than solely by theological innovation. This invites platform studies to move beyond content moderation toward re-evaluating how recommendation algorithms structure spiritual spaces. A central implication is the need for value-aligned algorithms that allow users to optimize for outcomes beyond engagement, such as contemplation or community connection.

\subsection{Limitations}

Our study has several limitations. First, our sampling strategy utilizes English-language queries sorted by view count; while capturing high-impact content, this may overrepresent algorithmically amplified videos and underrepresent authentic non-English traditions, such as Arabic Islamic or Sanskrit Hindu content. Given that the dataset spans over 130 languages, the inherent limitations of current cross-lingual translation and semantic alignment methods may compromise the accuracy and representativeness of comment analysis for non-English content. Second, although our LLM-assisted coding achieved robust reliability (Krippendorff's $\alpha$ 0.83--0.91), subtle tradition-specific symbolism may still present challenges despite systematic human validation. Third, the nascent adoption of AI-generated content resulted in a limited sample size (5.28\%, n=58), warranting cautious interpretation of associated effect sizes due to potential variance inflation. Finally, our cross-sectional design captures correlations rather than causality, and findings specific to YouTube may not fully generalize to platforms with different algorithmic logics. We acknowledge the risk of these taxonomies being used to optimize manipulative content and urge ethical adherence in future applications.

\section{Conclusion}

This study examines digital religious communication on YouTube through a mixed-methods analysis of 1,100 popular videos and 1.92 million comments across five major religions. Findings reveal that religious videos employ distinct multimodal strategies combining narrative frameworks with visual aesthetics, exhibiting both cross-religious templates and tradition-specific variations. Comment sections function as complex discourse ecologies preserving tradition-specific interaction patterns rather than homogenizing discourse. Technical production choices, particularly AI-generated content, show notable associations with engagement, revealing tensions between technological innovation and authenticity concerns. For content creators, our taxonomies illuminate connections between visual aesthetics, narrative strategies, and viewer responses. For platform designers, engagement patterns highlight the need for recommendation systems supporting contemplative engagement rather than optimizing solely for interaction metrics. For researchers, we provide a methodological framework combining LLM efficiency with human validation for large-scale analysis of religious media. Future work should explore algorithm redesigns accommodating diverse spiritual needs and AI tools respecting authenticity concerns central to religious communication.

\section*{Ethical Statement}

All data were collected from publicly accessible YouTube content via official API, in compliance with terms of service. No personally identifiable information was retained, and all analyses were conducted at the aggregate level. To minimize potential harm, we focus exclusively on high-engagement videos from established channels, thereby avoiding the inadvertent spotlighting of vulnerable individuals or small communities. Individual comments are not quoted verbatim in this paper. Raw comment data will not be publicly released, given the sensitive nature of personal experiences and beliefs contained therein. Raw video files are similarly unavailable for distribution due to copyright constraints. Only derived annotations and aggregate statistics will be made available to the research community, subject to platform policies.

\bibliography{aaai2026}

\section{Paper Checklist}

\begin{enumerate}

\item For most authors...
\begin{enumerate}
  \item Would answering this research question advance science without violating social contracts, such as violating privacy norms, perpetuating unfair profiling, exacerbating the socio-economic divide, or implying disrespect to societies or cultures?
    \answerTODO{Yes.}
    
  \item Do your main claims in the abstract and introduction accurately reflect the paper's contributions and scope?
    \answerTODO{Yes.}
    
  \item Do you clarify how the proposed methodological approach is appropriate for the claims made? 
    \answerTODO{Yes.}
    
  \item Do you clarify what possible artifacts are in the data used, given population-specific distributions?
    \answerTODO{Yes.}
    
  \item Did you describe the limitations of your work?
    \answerTODO{Yes.}
    
  \item Did you discuss any potential negative societal impacts of your work?
    \answerTODO{Yes.}
    
  \item Did you discuss any potential misuse of your work?
    \answerTODO{Yes.}
    
  \item Did you describe steps taken to prevent or mitigate potential negative outcomes of the research, such as data and model documentation, data anonymization, responsible release, access control, and the reproducibility of findings?
    \answerTODO{Yes.}
    
  \item Have you read the ethics review guidelines and ensured that your paper conforms to them?
    \answerTODO{Yes.}
\end{enumerate}

\item Additionally, if your study involves hypotheses testing...
\begin{enumerate}
  \item Did you clearly state the assumptions underlying all theoretical results?
    \answerTODO{Not applicable.}
    
  \item Have you provided justifications for all theoretical results?
    \answerTODO{Not applicable.}
    
  \item Did you discuss competing hypotheses or theories that might challenge or complement your theoretical results?
    \answerTODO{Not applicable.}
    
  \item Have you considered alternative mechanisms or explanations that might account for the same outcomes observed in your study?
    \answerTODO{Yes.}
    
  \item Did you address potential biases or limitations in your theoretical framework?
    \answerTODO{Yes.}
    
  \item Have you related your theoretical results to the existing literature in social science?
    \answerTODO{Yes.}
    
  \item Did you discuss the implications of your theoretical results for policy, practice, or further research in the social science domain?
    \answerTODO{Yes.}
\end{enumerate}

\item Additionally, if you are including theoretical proofs...
\begin{enumerate}
  \item Did you state the full set of assumptions of all theoretical results?
    \answerTODO{Not applicable.}
    
  \item Did you include complete proofs of all theoretical results?
    \answerTODO{Not applicable.}
\end{enumerate}

\item Additionally, if you ran machine learning experiments...
\begin{enumerate}
  \item Did you include the code, data, and instructions needed to reproduce the main experimental results (either in the supplemental material or as a URL)?
    \answerTODO{Not applicable.}
    
  \item Did you specify all the training details (e.g., data splits, hyperparameters, how they were chosen)?
    \answerTODO{Not applicable.}
    
  \item Did you report error bars (e.g., with respect to the random seed after running experiments multiple times)?
    \answerTODO{Not applicable.}
    
  \item Did you include the total amount of compute and the type of resources used (e.g., type of GPUs, internal cluster, or cloud provider)?
    \answerTODO{Not applicable.}
    
  \item Do you justify how the proposed evaluation is sufficient and appropriate to the claims made? 
    \answerTODO{Not applicable.}
    
  \item Do you discuss what is ``the cost'' of misclassification and fault (in)tolerance?
    \answerTODO{Not applicable.}
\end{enumerate}

\item Additionally, if you are using existing assets (e.g., code, data, models) or curating/releasing new assets, \textbf{without compromising anonymity}...
\begin{enumerate}
  \item If your work uses existing assets, did you cite the creators?
    \answerTODO{Yes.}
    
  \item Did you mention the license of the assets?
    \answerTODO{Yes.}
    
  \item Did you include any new assets in the supplemental material or as a URL?
    \answerTODO{No.}
    
  \item Did you discuss whether and how consent was obtained from people whose data you're using/curating?
    \answerTODO{Yes.}
    
  \item Did you discuss whether the data you are using/curating contains personally identifiable information or offensive content?
    \answerTODO{Yes.}
    
  \item If you are curating or releasing new datasets, did you discuss how you intend to make your datasets FAIR?
    \answerTODO{Not applicable.}
    
  \item If you are curating or releasing new datasets, did you create a Datasheet for the Dataset? 
    \answerTODO{Not applicable.}
\end{enumerate}

\item Additionally, if you used crowdsourcing or conducted research with human subjects, \textbf{without compromising anonymity}...
\begin{enumerate}
  \item Did you include the full text of instructions given to participants and screenshots?
    \answerTODO{Not applicable.}
    
  \item Did you describe any potential participant risks, with mentions of Institutional Review Board (IRB) approvals?
    \answerTODO{Not applicable.}
    
  \item Did you include the estimated hourly wage paid to participants and the total amount spent on participant compensation?
    \answerTODO{Not applicable.}
    
  \item Did you discuss how data is stored, shared, and deidentified?
    \answerTODO{Yes. All data are stored securely and analyzed at aggregate levels without identifiers.}
\end{enumerate}

\end{enumerate}

\clearpage
\appendix
\onecolumn
\section{Appendix}
\label{sec:appendix}

\subsection{Statistical Tables and Figures}

\begin{table}[htbp]
  \centering
  \label{tab:A1_video_statistics}
  \begin{tabular}{@{}llrrrrr@{}}
    \toprule
    \textbf{Category} & \textbf{Metric} & \textbf{Mean} & \textbf{Median} & \textbf{Std Dev} & \textbf{Min} & \textbf{Max} \\
    \midrule
    \multirow{4}{*}{Overall} & Views & 5,060,496 & 965,428 & 14,975,658 & 3,937 & 255,550,421 \\
                            & Likes & 93,250 & 25,096 & 223,735 & 165 & 2,767,373 \\
                            & Comments & 2,623 & 1,100 & 3,543 & 101 & 19,000 \\
                            & Duration (s) & 307 & 60 & 406 & 4 & 1,788 \\
    \midrule
    \multirow{4}{*}{Short} & Views & 3,174,938 & 594,788 & 9,634,636 & 5,636 & 144,311,925 \\
                          & Likes & 113,259 & 27,952 & 264,231 & 165 & 2,767,373 \\
                          & Comments & 2,342 & 923 & 3,416 & 103 & 19,000 \\
                          & Duration (s) & 89 & 41 & 210 & 4 & 1,657 \\
    \midrule
    \multirow{4}{*}{Normal} & Views & 6,952,922 & 1,461,470 & 18,693,336 & 3,937 & 255,550,421 \\
                           & Likes & 73,168 & 21,386 & 171,783 & 191 & 2,389,170 \\
                           & Comments & 2,905 & 1,400 & 3,649 & 101 & 19,000 \\
                           & Duration (s) & 526 & 391 & 436 & 5 & 1,788 \\
    \bottomrule
  \end{tabular}
  \caption{Descriptive Statistics of Video Dataset (N=1,100)}
\end{table}

\begin{table}[htbp]
\centering
\label{tab:A2_descriptive_stats_revised}
\begin{tabular}{lrrrrr}
\toprule
 & Min & Max & Median & Mean & Std Dev \\
\midrule
Subscribers & 377 & 293,000,000 & 137,000 & 1,532,333.46 & 11,085,867.59\\
Views & 20,030 & 296,016,464,229 & 35,201,025.5 & 825,509,599.50 & 10,571,056,170.21 \\
Videos & 1 & 235,424 & 330.5 & 1,915.29 & 11,968.94\\
\bottomrule
\end{tabular}
\caption{Descriptive Statistics of Creator Channels (N=820)}
\end{table}

\begin{table}[htbp]
\centering
\begin{threeparttable}
\label{tab:A3_content_strategy_en}
\begin{tabular}{l@{\hspace{0.3em}}l@{\hspace{0.3em}}r@{\hspace{0.3em}}r@{\hspace{0.5em}}rrrrr}
\toprule
\begin{tabular}[t]{@{}p{1.8cm}@{}}\textbf{Main}\\\textbf{Dimension}\end{tabular} & \textbf{Subcategory} & \textbf{Overall} & \textbf{S/N Diff} & \multicolumn{5}{c}{\textbf{Distribution by Religion (\%)}} \\[-9pt]
& & \textbf{(\%)} & \textbf{(\%)} & & & & & \\
\cmidrule(lr){5-9}
& & & & \textbf{Buddhism} & \textbf{Christianity} & \textbf{Hinduism} & \textbf{Islam} & \textbf{Judaism} \\
\midrule
\multirow{5}{*}{\begin{tabular}[t]{@{}p{1.8cm}@{}}\textbf{Video}\\\textbf{Type}\end{tabular}} & Narrative/Speaking & 62.09 & +5.41 & 70.32 & 59.82 & 57.92 & 65.40 & 57.33 \\
& Event Recording & 17.27 & +6.85 & 17.81 & 11.16 & 23.98 & 17.06 & 16.44 \\
& Music & 12.45 & -16.96 & 2.28 & 25.45 & 6.79 & 7.11 & 20.00 \\
& Other & 4.82 & +0.16 & 8.68 & 0.89 & 6.79 & 6.16 & 1.78 \\
& Dance & 3.36 & +4.54 & 0.91 & 2.68 & 4.52 & 4.27 & 4.44 \\
\midrule
\multirow{7}{*}{\begin{tabular}[t]{@{}p{1.8cm}@{}}\textbf{Speech}\\\textbf{Strategy}\end{tabular}} & Teaching \& Guidance & 34.73 & -0.85 & 47.49 & 25.89 & 33.48 & 31.28 & 35.56 \\
& Other & 23.64 & -2.27 & 23.74 & 19.64 & 28.51 & 18.48 & 27.56 \\
& Interview \& Dialogue & 11.18 & -1.31 & 4.11 & 9.82 & 13.12 & 16.11 & 12.89 \\
& Stories \& Parables & 10.91 & -0.76 & 13.24 & 11.61 & 14.03 & 9.48 & 6.22 \\
& Preaching \& Proclamation & 9.00 & -6.76 & 7.76 & 16.52 & 6.33 & 9.95 & 4.44 \\
& Debate \& Apologetics & 6.45 & +7.79 & 2.28 & 11.61 & 2.26 & 9.00 & 7.11 \\
& Testimony \& Sharing & 4.09 & +4.17 & 1.37 & 4.91 & 2.26 & 5.69 & 6.22 \\
\midrule
\multirow{5}{*}{\begin{tabular}[t]{@{}p{1.8cm}@{}}\textbf{Religious}\\\textbf{Narrative}\end{tabular}} & Salvation Narrative & 46.36 & -14.35 & 45.21 & 63.84 & 34.84 & 46.45 & 41.33 \\
& Other & 30.09 & +1.16 & 16.89 & 24.11 & 38.46 & 31.75 & 39.11 \\
& Progress Narrative & 9.55 & +1.97 & 20.55 & 2.68 & 14.03 & 6.16 & 4.44 \\
& Harmony Narrative & 8.36 & +3.60 & 15.98 & 2.68 & 8.14 & 6.16 & 8.89 \\
& Conflict Narrative & 5.64 & +7.62 & 1.37 & 6.70 & 4.52 & 9.48 & 6.22 \\
\midrule
\multirow{7}{*}{\begin{tabular}[t]{@{}p{1.8cm}@{}}\textbf{Persuasion}\\\textbf{Strategy}\end{tabular}} & Authority Dependence & 42.27 & -10.52 & 43.38 & 44.64 & 35.29 & 47.87 & 40.44 \\
& Experiential Evidence & 34.36 & +8.97 & 31.96 & 35.71 & 35.29 & 35.55 & 33.33 \\
& Other & 12.45 & -2.41 & 9.13 & 10.27 & 19.00 & 8.06 & 15.56 \\
& Logical Argument & 5.36 & +4.16 & 6.39 & 4.91 & 4.98 & 4.27 & 6.22 \\
& Emotional Resonance & 3.36 & -0.93 & 7.76 & 3.57 & 1.36 & 3.32 & 0.89 \\
& Scientific Support & 1.45 & +0.72 & 0.91 & 0.89 & 3.62 & 0.95 & 0.89 \\
& Identity Identification & 0.73 & 0.00 & 0.46 & 0.00 & 0.45 & 0.00 & 2.67 \\
\bottomrule
\end{tabular}
\begin{tablenotes}
  \item S/N Diff = Short/Normal Video Difference, indicating the percentage difference between short-form and normal-length Video content strategies.
\end{tablenotes}
\end{threeparttable}
\caption{Distribution of Narrative Content Strategies across Religions (\%)}
\end{table}

\begin{table}[htbp]
\centering
\begin{threeparttable}
\label{tab:A4_sacred_items_en}
\begin{tabular}{l@{\hspace{0.3em}}l@{\hspace{0.3em}}r@{\hspace{0.3em}}r@{\hspace{0.5em}}rrrrr}
\toprule
\begin{tabular}[t]{@{}p{1.8cm}@{}}\textbf{Main}\\\textbf{Dimension}\end{tabular} & \textbf{Subcategory} & \textbf{Overall} & \textbf{S/N Diff}\tnote{a} & \multicolumn{5}{c}{\textbf{Distribution by Religion (\%)}} \\[-9pt]
& & \textbf{(\%)} & \textbf{(\%)} & & & & & \\
\cmidrule(lr){5-9}
& & & & \textbf{Buddhism} & \textbf{Christianity} & \textbf{Hinduism} & \textbf{Islam} & \textbf{Judaism} \\
\midrule
\multirow{5}{*}{\begin{tabular}[t]{@{}p{1.8cm}@{}}\textbf{Sacred}\\\textbf{Items}\end{tabular}} & No Sacred Items & 60.18 & +2.32 & 41.55 & 72.32 & 42.99 & 72.99 & 71.11 \\
& Mixed Types & 18.09 & -6.43 & 24.20 & 8.93 & 36.65 & 9.48 & 11.11 \\
& Religious Books/Scripture & 12.36 & -0.77 & 19.63 & 13.84 & 5.43 & 10.43 & 12.44 \\
& Sacred Objects & 5.27 & +1.80 & 9.13 & 1.79 & 8.60 & 3.32 & 3.56 \\
& Ritual Items & 4.09 & +3.08 & 5.48 & 3.12 & 6.33 & 3.79 & 1.78 \\
\bottomrule
\end{tabular}
\begin{tablenotes}
 \item[a] S/N Diff = Short/Normal Video Difference, indicating the percentage difference between short-form and normal-length video content strategies.
\end{tablenotes}
\end{threeparttable}
\caption{Sacred Items Statistics}
\end{table}

\begin{table}[htbp]
\centering
\small
\begin{threeparttable}
\label{tab:A5_comment_strategy_en}
\begin{tabular}{
  l@{\hspace{0.2em}}
  >{\raggedright\arraybackslash}p{3.2cm}
  @{\hspace{0.2em}}r
  @{\hspace{0.2em}}r
  @{\hspace{0.4em}}rrrrr
}
\toprule
\begin{tabular}[t]{@{}p{2.0cm}@{}}
\textbf{Main}\\\textbf{Dimension}
\end{tabular}
& \textbf{Subcategory}
& \textbf{Overall}
& \textbf{S/N Diff}
& \multicolumn{5}{c}{\textbf{\shortstack{Distribution by \\ Religion (\%)}}} \\[-9pt]
& 
& \textbf{(\%)}
& \textbf{(\%)}
& 
& 
& 
& 
& \\
\cmidrule(lr){5-9}
& 
& 
& 
& \textbf{Buddhism}
& \textbf{Christianity}
& \textbf{Hinduism}
& \textbf{Islam}
& \textbf{Judaism} \\
\midrule
\multirow{6}{*}{\begin{tabular}[t]{@{}p{2.0cm}@{}}\textbf{Expression}\\\textbf{Pattern (L1)}\end{tabular}}
& 1. Stance & 54.11 & +3.67 & 54.19 & 60.25 & 42.33 & 57.01 & 48.90 \\
& 2. Questioning & 9.71 & +3.78 & 8.97 & 8.69 & 9.33 & 9.91 & 13.31 \\
& 3. Sharing & 15.07 & -3.55 & 17.22 & 16.08 & 11.81 & 11.57 & 15.74 \\
& 4. Seeking Interaction & 1.86 & -0.24 & 2.44 & 1.55 & 1.42 & 1.60 & 1.79 \\
& 5. Call to Action & 0.82 & +0.29 & 1.05 & 0.55 & 0.57 & 0.77 & 1.05 \\
& Other & 18.43 & -3.95 & 16.14 & 12.89 & 34.54 & 19.13 & 19.21 \\
\midrule
\multirow{12}{*}{\begin{tabular}[t]{@{}p{2.0cm}@{}}\textbf{Expression}\\\textbf{Pattern (L2)}\end{tabular}}
& 1.1 Agreement \& Approval & 43.69 & -2.40 & 44.63 & 52.08 & 31.37 & 44.70 & 35.48 \\
& 1.2 Opposition \& Denial & 7.19 & +4.32 & 6.65 & 5.56 & 7.11 & 8.54 & 9.59 \\
& 1.3 Criticism \& Satire & 1.92 & +1.27 & 1.53 & 1.43 & 2.44 & 2.41 & 2.63 \\
& 2.1 Factual Questioning & 5.29 & +1.79 & 4.14 & 5.02 & 5.56 & 5.75 & 7.65 \\
& 2.2 Logical Rebuttal & 2.27 & +0.78 & 2.04 & 2.21 & 2.19 & 2.28 & 2.97 \\
& 2.3 Stance Questioning & 1.49 & +1.02 & 2.26 & 0.87 & 0.82 & 1.21 & 1.67 \\
& 3.1 Experience \& Testimony & 6.63 & -4.66 & 6.06 & 9.46 & 5.64 & 4.79 & 6.36 \\
& 3.2 Knowledge Sharing & 3.55 & +1.90 & 5.21 & 2.15 & 2.51 & 3.03 & 3.68 \\
& 3.3 Rec. \& Guidance & 0.90 & -0.95 & 0.53 & 1.16 & 0.89 & 0.92 & 1.32 \\
& 3.4 Expressing Opinion & 3.60 & +0.04 & 4.76 & 3.08 & 2.62 & 2.51 & 4.12 \\
& 4.1 Seeking Dialogue & 1.04 & +0.32 & 1.81 & 0.61 & 0.26 & 0.86 & 0.85 \\
& Other & 22.42 & -3.43 & 20.38 & 16.38 & 38.58 & 23.00 & 23.69 \\
\midrule
\multirow{11}{*}{\begin{tabular}[t]{@{}p{2.0cm}@{}}\textbf{Comment}\\\textbf{Topic}\end{tabular}}
& 1. Religious Figures & 17.65 & +7.25 & 18.23 & 16.57 & 26.26 & 16.75 & 12.17 \\
& 2. Doctrine \& Creed & 19.55 & -0.18 & 18.95 & 24.79 & 11.26 & 19.75 & 18.36 \\
& 3. Religious Stories & 2.56 & +0.15 & 2.59 & 3.10 & 1.97 & 1.73 & 3.17 \\
& 4. Religious Practice & 6.68 & +0.38 & 9.46 & 5.01 & 2.77 & 8.44 & 3.81 \\
& 5. Comparative Religion & 1.75 & +1.22 & 2.10 & 1.01 & 1.70 & 1.93 & 1.99 \\
& 6. Video Feedback & 9.90 & -0.58 & 12.17 & 10.92 & 6.18 & 7.88 & 8.62 \\
& 7. Social Issues & 4.71 & +1.61 & 6.14 & 2.47 & 2.33 & 3.72 & 8.64 \\
& 8. Science & 0.38 & +0.34 & 0.56 & 0.14 & 0.30 & 0.50 & 0.27 \\
& 9. Pop Culture & 4.12 & -0.40 & 6.61 & 3.28 & 0.78 & 3.30 & 3.59 \\
& 10. Economic Topics & 0.09 & +0.02 & 0.13 & 0.03 & 0.05 & 0.06 & 0.20 \\
& Other Topics & 32.61 & -9.81 & 23.06 & 32.68 & 46.39 & 35.95 & 39.18 \\
\midrule
\multirow{8}{*}{\begin{tabular}[t]{@{}p{2.0cm}@{}}\textbf{Emotional}\\\textbf{Expression}\end{tabular}}
& 1. Approval & 25.52 & -10.08 & 25.65 & 27.73 & 19.47 & 23.57 & 29.10 \\
& 2. Anger & 9.24 & +3.82 & 7.57 & 7.50 & 11.44 & 10.16 & 13.16 \\
& 3. Sarcasm & 7.43 & +4.98 & 6.66 & 5.29 & 9.26 & 9.15 & 9.14 \\
& 4. Confusion & 7.49 & +0.65 & 6.90 & 7.06 & 9.05 & 7.61 & 8.17 \\
& 5. Calmness & 19.09 & +0.90 & 22.00 & 16.16 & 19.97 & 18.18 & 17.82 \\
& 6. Awe & 26.12 & -1.66 & 24.55 & 33.12 & 24.56 & 26.54 & 18.15 \\
& 7. Surprise & 2.07 & +0.19 & 2.62 & 1.77 & 1.67 & 1.84 & 1.93 \\
& 8. Other & 3.06 & +1.19 & 4.05 & 1.38 & 4.59 & 2.93 & 2.52 \\
\bottomrule
\end{tabular}
\begin{tablenotes}
  \item[a] S/N Diff = Short/Normal Video Difference, indicating the percentage difference between comment strategies on short-form versus normal-length videos.
\end{tablenotes}
\end{threeparttable}
\caption{Distribution of Comment Strategies across Religions (\%)}
\end{table}

\begin{table}[htbp]
\centering
\begin{threeparttable}
\label{tab:A6_correlation_analysis_en}
\begin{tabular}{l@{\hspace{1em}}l@{\hspace{1em}}r@{\hspace{1em}}r@{\hspace{1em}}r@{\hspace{1em}}r}
\toprule
\textbf{Main Dimension} & \textbf{Independent Variable} & \textbf{$\chi^2$} & \textbf{p-value} & \textbf{df} & \textbf{Cramér's V} \\
\midrule
\multirow{7}{*}{\begin{tabular}[t]{@{}p{2.2cm}@{}}\textbf{Expression}\\\textbf{Pattern (L1)}\end{tabular}} 
& AI-generated content & 4926.45 & $<.001$ & 6 & 0.387 \\
& Video Type & 54572.44 & $<.001$ & 18 & 0.157 \\
& Color Tone & 29637.55 & $<.001$ & 12 & 0.147 \\
& Speech Strategy & 89041.19 & $<.001$ & 36 & 0.140 \\
& Sacred Objects & 3730.17 & $<.001$ & 12 & 0.131 \\
& Religious Narrative Framework & 47523.94 & $<.001$ & 24 & 0.126 \\
& Persuasion Strategy & 61731.32 & $<.001$ & 36 & 0.117 \\
\midrule
\multirow{9}{*}{\begin{tabular}[t]{@{}p{2.2cm}@{}}\textbf{Expression}\\\textbf{Pattern (L2)}\end{tabular}}
& AI-generated content & 5514.33 & $<.001$ & 9 & 0.410 \\
& Religious Narrative Framework & 58993.03 & $<.001$ & 27 & 0.191 \\
& Speech Strategy & 58211.62 & $<.001$ & 36 & 0.177 \\
& Video Type & 61025.97 & $<.001$ & 27 & 0.166 \\
& Color Tone & 35467.12 & $<.001$ & 18 & 0.161 \\
& Sacred Objects & 4209.79 & $<.001$ & 18 & 0.139 \\
& Shot Type & 20187.12 & $<.001$ & 27 & 0.137 \\
& Persuasion Strategy & 66882.81 & $<.001$ & 54 & 0.122 \\
& B-Roll & 7652.21 & $<.001$ & 9 & 0.101 \\
\midrule
\multirow{8}{*}{\begin{tabular}[t]{@{}p{2.2cm}@{}}\textbf{Comment}\\\textbf{Topic}\end{tabular}}
& AI-generated content & 6601.63 & $<.001$ & 6 & 0.448 \\
& Sacred Items & 8045.09 & $<.001$ & 12 & 0.192 \\
& Video Type & 41246.02 & $<.001$ & 18 & 0.136 \\
& Shot Type & 18766.80 & $<.001$ & 18 & 0.132 \\
& Religious Narrative Framework & 21067.28 & $<.001$ & 18 & 0.114 \\
& Speech Strategy & 55147.96 & $<.001$ & 36 & 0.110 \\
& Symbol Display & 19912.80 & $<.001$ & 18 & 0.108 \\
& Color Tone & 15039.38 & $<.001$ & 12 & 0.105 \\
\midrule
\multirow{9}{*}{\begin{tabular}[t]{@{}p{2.2cm}@{}}\textbf{Emotional}\\\textbf{Expression}\end{tabular}}
& AI-generated content & 1451.61 & $<.001$ & 7 & 0.210 \\
& Religious Narrative Framework & 64174.29 & $<.001$ & 21 & 0.199 \\
& Speech Strategy & 64781.01 & $<.001$ & 28 & 0.186 \\
& Video Type & 49717.26 & $<.001$ & 21 & 0.150 \\
& Sacred Objects & 4822.67 & $<.001$ & 14 & 0.149 \\
& Color Tone & 26095.87 & $<.001$ & 14 & 0.138 \\
& Shot Type & 19267.79 & $<.001$ & 21 & 0.133 \\
& Persuasion Strategy & 55473.91 & $<.001$ & 42 & 0.111 \\
& B-Roll & 8135.95 & $<.001$ & 7 & 0.104 \\
\bottomrule
\end{tabular}

\end{threeparttable}
\caption{Correlation Analysis of Video Features and Comment Strategies (Cramér's V $> 0.1$)}
\end{table}

\begin{figure}[htbp]
  \centering
  \includegraphics[width=0.9\textwidth]{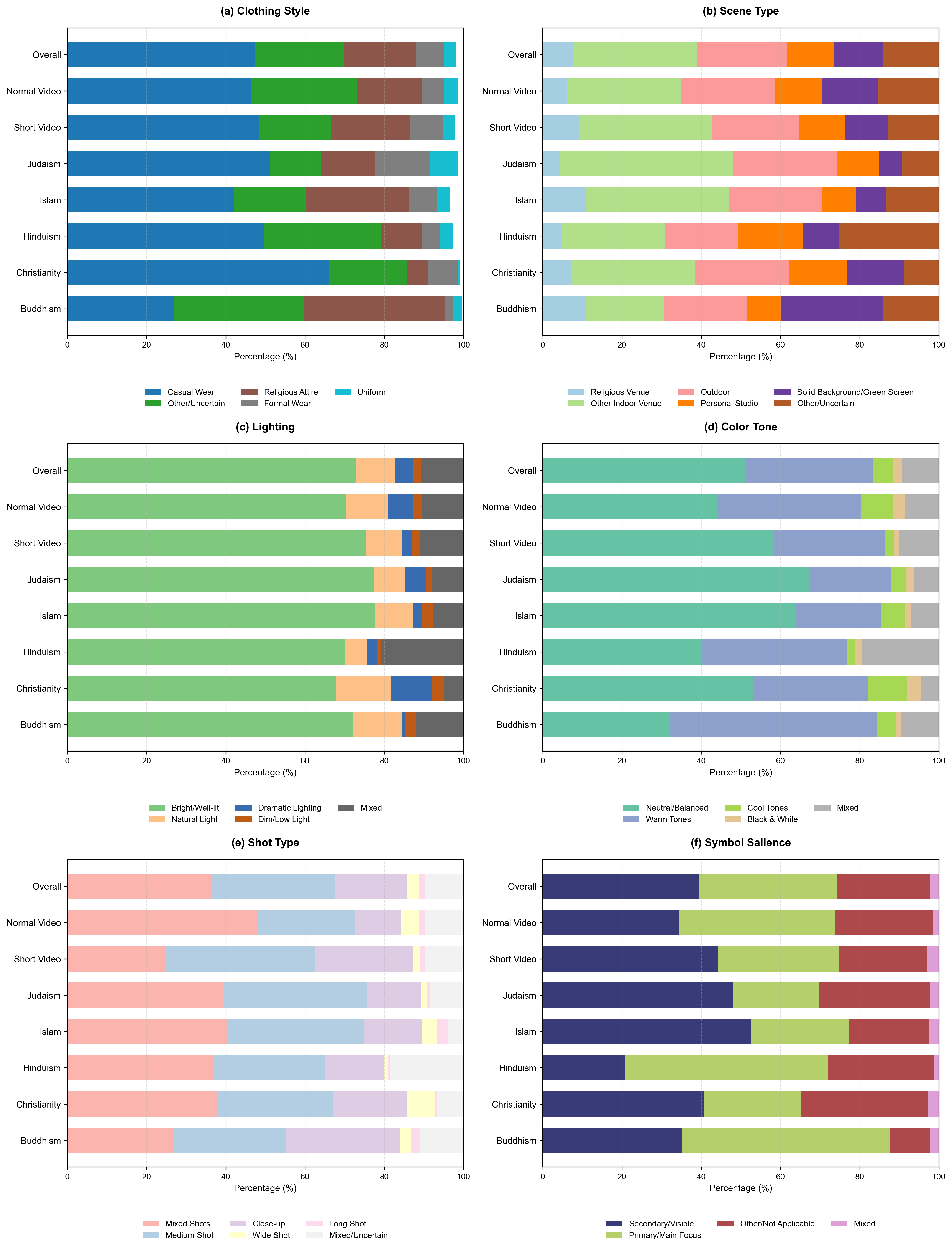}
\caption{Distribution of Visual Features in Religious Videos.}
  \label{fig:A1_broll}
\end{figure}

\end{document}